\documentclass[12pt]{article}
\usepackage{amssymb}

\usepackage{amsmath}

\begin{document}

\begin{center}
\vspace{1cm}{\Large {\bf Little Groups and Statistics of Branes}}

\vspace{1cm} {\bf R.Mkrtchyan} \footnote{ E-mail: mrl@r.am}
\vspace{1cm}

\vspace{1cm}

{\it Theoretical Physics Department,} {\it Yerevan Physics
Institute}

{\it Alikhanian Br. St.2, Yerevan, 375036 Armenia }
\end{center}

\vspace{1cm}
\begin{abstract}
The little groups (i.e. the subgroups of Lorentz group, leaving
invariant given configurations of tensorial charges) of unitary
irreps of superstring/M-theory superalgebras are considered. It is
noted, that in the case of $(n-1)/n$ (maximal supersymmetric) BPS
configuration in any dimensions the non-zero supercharge is
neutral w.r.t. the algebra of little group, which means that all
members of supermultiplet are in the same representation of that
algebra and hence of (generalized with tensorial charges) Poincare
algebra. This situation is similar to two-dimensional case and
shows that usual spin-statistics connection statement is
insufficient in the presence of branes, because different little
groups can appear. We discuss the rules for definition of
statistics for representations of generalized Poincare, and note
that a geometric quantization method seems to be most relevant for
that purpose.

\end{abstract}

\renewcommand{\thefootnote}{\arabic{footnote}} \setcounter{footnote}0
{\smallskip \pagebreak }

\section{Introduction}

    Many features of modern superstring theories can be deduced
directly from considerations of their supersymmetry algebras
\cite{F}. Most general among them is the M-theory  $N=1, d=11$
superalgebra
\begin{eqnarray}
\{ Q,\bar Q\}  &=& \gamma ^\mu  P_\mu + \gamma ^{\mu \nu } Z_{\mu
\nu }  + \gamma ^{\mu \nu \lambda \rho \sigma } Z_{\mu \nu
\lambda \rho \sigma },  \label{1} \\
\mu,\nu,... &=&0,1,2,..10.  \notag
\end{eqnarray}
(plus relations, including Lorentz generators) where Q is a
Majorana spinor. We shall consider also the simpler analog in
$d=4$, the $N=1$ supersymmetry algebra

\begin{eqnarray}
\{ Q,\bar Q\}  &=& \gamma ^\mu  P_\mu + \frac {1}{2} \gamma ^{\mu
\nu } Z_{\mu
\nu },  \label{2} \\
\mu,\nu,... &=&0,1,2,3.\notag
\end{eqnarray}
with Majorana $Q$.

Let's construct one of the simplest ("particle") representation of
(\ref{2}), i.e. the representation for which the vector $P_{\mu}$
is non-zero, and all tensorial charges are zero. At first steps of
construction of (unitary) irreps for such an algebras one have to
fix the values of all Casimirs, constructed from $P_{\mu},
Z_{\mu\nu}, ...$, and take the particular point on that orbit.
Let's take the point $P_{\mu}=(m,0,0,...), Z=0$. The stabilizer
(i.e. little group) of this point on the orbit  is SO(3). The
algebra (\ref{2}) becomes, in a two-component notations (of e.g.
\cite{bag}:
\begin{eqnarray}
\begin{gathered}
  \{ Q_A ,
\bar Q_{\dot B} \}  = m \delta_{A\dot B}  \hfill \\
  \{ Q_A ,Q_B \}  = 0  \hfill \\
  \{ \bar Q_{\dot A} ,\bar Q_{\dot B} \}  = 0 \hfill \\
\end{gathered}  \label{3}
\end{eqnarray}
where $\bar Q_ {\dot A}$ is a Hermitian conjugate to $Q_{\dot A}$,
so one of them can be considered as creation, and second one as
annihilation  operators. The representation of an algebra
(\ref{3}) can be constructed by taking a "vacuum" $ |s>$ in any
unitary representation of SO(3) with spin $s$, which is
annihilated by operators $Q$, then applying the creation operators
$\bar Q$ as many times as possible, and finally inducing the
representation to the whole super-Poincare group. So the whole
supermultiplet before induction will be a collection of few
irreducible representations of SO(3), transforming one into
another under an action of $Q, \bar Q$. More exactly, $Q, \bar Q$
will transform states with integer spins into those with
half-integer and vice-versa, because $Q, \bar Q$ itself have spin
one-half. This is in agreement with spin-statistics connection at
$d=4$, because $Q,{\bar Q}$ are fermionic and flip the statistics.

Algebra (\ref{3}) is an algebra of $2$ pairs of fermionic
creation-annihilation operators, so the number of states in the
supermultiplet described is maximal: $2^2$, composed of $2$
fermions plus $2$ bosons.

The number of states becomes less in the so called shortened (BPS)
supermultiplets, playing an important role in these theories.
Their existence is based on a specific features of (\ref{1}),
(\ref{2}) and similar superalgebras. For our model example
(\ref{2}) such a multiplet appears e.g. for $P_{\mu}=(1,0,0,1),
Z_{\mu\nu}=0$, which corresponds to massless particle. Then
(\ref{2}) becomes

\begin{eqnarray}
\begin{gathered}
  \{ Q_1 ,\bar Q_{\dot 1} \}  = 2  \hfill \\
   \{ Q_2 ,\bar Q_{\dot 2} \}  = 0  \hfill \\
  \{ Q_A ,Q_B \}  = 0  \hfill \\
  \{ \bar Q_{\dot A} ,\bar Q_{\dot B} \}  = 0  \hfill \label{33}\\
\end{gathered}
\end{eqnarray}

The little group for this massless particle case is a semidirect
product of $SO(2)$ on a group of two-dimensional translation, i.e.
it is a two-dimensional Euclidean Poincare. Supercharge $Q_{1}$
has spin (helicity) 1/2 w.r.t. the SO(2) (which acts as
multiplication by a phase factor on $Q_{1}$), so it is
transforming half-integer helicity states into integer ones and
vise-versa. Due to relations (\ref {33}) the half of components of
operator $\bar Q$ are represented by zero, so there is only one
creation operator, the number of states is $1+1$, half of them
fermionic, with half-integer helicity and another half bosonic,
with integer helicity.

Another BPS multiplet appears for BPS membrane, i.e. the charges
configuration $P_{\mu}=(m,0,0,0)$, $Z_{12}=-Z_{21}=m$, other
components of $Z_{\mu\nu}$  zero. Situation is similar to that of
massless particle.  The little group evidently is an $SO(2)$ group
of rotations around $z$ axis. Non-zero $Q$ have a spin (helicity)
$\pm \frac{1}{2}$ w.r.t. this little group, so states connected by
$Q$ have helicities, differing by {1/2}. These facts can be
deduced from the main relation (\ref{2}), which in this case
acquire a form (in two-component notations, only non-zero
relations are written):

\begin{eqnarray}
\begin{gathered}
  \{ Q_A ,\bar Q_{\dot B} \}  = m\delta_{A\dot B}  \hfill \\
  \{ Q_1 ,Q_2 \}  = m i  \hfill \\
  \{ \bar Q_{\dot 1} ,\bar Q_{\dot 2} \}  = -mi  \hfill \label{44}\\
\end{gathered}
\end{eqnarray}

The  little group of rotation around $z$ axis, with generator
$\sigma^{12}$ (in notations of \cite{bag}), acts on a $Q$ spinors
as multiplication on a phase factor for $Q_1$ and opposite phase
factor multiplication for $Q_2$. Let's introduce a combinations

\begin{eqnarray}
\begin{gathered}
 a= Q_2+i\bar Q_{\dot 1}\\
 \bar a= \bar Q_{\dot 2}-i Q_1 \label{55}\\
b= Q_2-i\bar Q_{\dot 1}\\
 \bar b= \bar Q_{\dot 2}+i Q_1
\end{gathered}
\end{eqnarray}
It is easy to check using (\ref{44}) that operators $b, \bar b$
anticommute with all $Q$ operators, and hence will be represented
by zero matrixes. Operators $a, \bar a$ are usual fermionic
creation-annihilation operators. Dimensionality of representation
is 1+1. According to abovementioned properties of $Q$ under little
group rotations, we see that operators $a,  \bar a$ have a
helicity $\pm 1/2$.

 Taking different "vacuums" $|s>$ with spin $s$
we shall obtain an irreps of susy algebra (\ref{55}), which can be
called "BPS membranes with spins". (One may ask for a Lagrangians
for that branes. Recall that in similar situation for particles
with spins one has to introduce an internal degrees of freedom,
e.g. internal sphere $S^2$. One can expect something similar here,
although we are not aware of any such considerations.
Particularly, we don't know classical solutions for
supergravities, which can be identified with "branes with spin".)

 The problem can be formulated at this stage. The usual
spin-statistics theorem is derived for usual $d=4$ Poincare
algebra, and claims that states with integer spins have a Bose
statistics, and states with half-integer spins - Fermi one. In
higher dimensions spin (or helicity)is substituted by some
representation of little groups, which are of $SO$ type (more
exactly, $SO$ groups are their compact part), and spin-statistics
statement is that spin-tensor representations are fermionic and
purely tensorial representations - of bosonic type. This can be
deduced, particularly, by  dimensional reduction, assuming that
statistics is not being changed by that process. This is enough to
describe spin-statistics connection for usual Poincare
(super)algebras. The problem is that in the presence of brane
charges the little groups, and representations of $Q$ with respect
to these little groups can be different. In that case one have to
generalize the statement of spin-statistics connection. So, the
problem is: for all brane superalgebras  (\ref{1}), (\ref{2}),
etc., find out, for all (physical) orbits, the corresponding
little groups and representations of $Q$ with respect to this
little groups. Then one has to assign Fermi or Bose or both
statistics to each of these representations. Actually situation
certainly will be different in comparison with standard at $d=4$,
in that it is possible, that the same representation of Poincare
algebra can have both types of statistics, as we shall see below.
This resembles the $d=2$ situation, when there is no spin, and the
same Poincare algebra's representations can have both statistics.
The assignment of Fermi-Bose statistics has to be in agreement
with the fermionic nature of $Q$, i.e. the fact that it is
changing statistics. It is not clear now what kinds of little
groups are possible in different dimensions.

\section{N=1, d=4, 3/4 BPS Supermultiplet}
Now we shall present a promised example at $d=4$, in which usual
spin-statistics connection is not applicable. That is the BPS
representation, maintaining maximal $(n-1)/n$ supersymmetry, first
described at $d=4$ in \cite{band}, and called preons in
\cite{band2}. Let's take $P_{\mu}$ and $Z_{\mu\nu}$ such that
$(\Gamma ^{i}P_{i}+\Gamma ^{ij}Z_{ij})_{\alpha\beta}=\lambda
_\alpha \lambda _\beta $. Then (\ref{2}) becomes

\begin{eqnarray}
\left\{ {Q_\alpha},Q_\beta \right\} &=&\lambda _\alpha  \lambda _\beta  \label{66} \\
\alpha,\beta,... &=&1,2,3,4.  \notag
\end{eqnarray}

This configuration satisfies positivity restriction, i.e. the
eigenvalues of r.h.s. matrix are non-negative. For this orbit the
algebra of little group is $T^2$, i.e. two-dimensional Abelian
subalgebra of Lorentz algebra. We are interested what are the
representations of non-zero $Q$-s w.r.t this algebra. Assuming
that only first component of $\lambda _\alpha $ is non-zero we can
construct the minimal representation of (\ref{66}), by
representing all component of $Q$ except first one by zero, and
first one by two by two matrix. Then it is evident that Lorentz
generators which leave $\lambda _\alpha$ invariant, will leave
invariant $Q_\alpha$, also. Explicitly, in  two-component
notations, these statements look as follows. Anticommutators are
\begin{eqnarray}
\begin{gathered}
  \{ Q_A ,
\bar Q_{\dot B} \}  = \lambda_{A} \lambda_{\dot B}   \\
  \{ Q_A ,Q_B \}  = \lambda_{A} \lambda _{B}   \\
  \{ \bar Q_{\dot A} ,\bar Q_{\dot B} \}  = \bar \lambda_{\dot
  A} \bar \lambda_{\dot B} \hfill \\
\end{gathered}  \label{77}
\end{eqnarray}
Take for definiteness $\lambda_{A}=(1,0)$, then little algebra has
two non-zero elements. Let $\omega_{\mu\nu}M_{\mu\nu}$ be the
general element of Lorentz algebra, where $M_{\mu\nu}$ are
generators. Then one element of little group has non-zero
parameters $\omega_{01}=\omega_{13}=-\omega_{10}=-\omega_{31}$
only, another one
$\omega_{02}=\omega_{23}=-\omega_{20}=-\omega_{312}$ only. These
two elements are commuting with each other. So, little algebra is
a two-dimensional Abelian algebra, which acts trivially on
$\lambda_{A}=(1,0)$. Algebra (\ref{77}) becomes (non-zero
anticommutators only):
\begin{eqnarray}
\begin{gathered}
  \{ Q_1 ,
\bar Q_{\dot 1} \}  = 1   \\
  \{ Q_1 ,Q_1 \}  = 1   \\
  \{ \bar Q_{\dot 1} ,\bar Q_{\dot 1} \}  = 1 \\
\end{gathered}  \label{77}
\end{eqnarray}
Introducing combinations $b=(Q_1 + \bar Q_{\dot 1})/\sqrt 2$ and
$b_{1}=(Q_1 -\bar Q_{\dot 1})/\sqrt 2$, we see that $b_1$ commute
with all $Q$-s and should be represented by zero matrix. Hermitian
operator $b$ satisfies a relation
\begin{eqnarray}
\begin{gathered}
  \{ b , b \}  = 2   \\
\end{gathered}
\end{eqnarray}
This algebra can be represented in two-dimensional space by taking
$b$ as two by two antidiagonal matrix with non-zero elements equal
to $1$. The dimensionality of representation is $2=1+1$. The
difference with above representations for e.g. massless particle,
is in that in this case we can choose representation real, with
real dimensionality $1+1$. This is possible because operator $b$
is Hermitian and doesn't change phase under little group
transformation, as is the case for operators $a, \bar a$ for
massless particle. So, as expected, in this case with $3$
supersymmetries surviving \cite{band}, the supermultiplet becomes
even shorter than for massless particle or membrane.

Thus, for the orbit considered, the non-zero components of
supercharge are neutral w.r.t. the corresponding little algebras.
In other words, for any representation $|s>$ the $b|s>$ is the
same representation of the little algebra. So, $b$ is not changing
any "spin", but is changing the statistics, due to its fermionic
nature. Evidently, this situation is similar to two dimensional
one for usual Poincare algebra: both fermions and bosons are
realizing the same representation.

\section{Conclusion: Higher Dimensions, Geometric Quantization}

These considerations evidently are applicable for all dimensions,
provided r.h.s. of anticommutator of supercharges is of rank one,
i.e. equal to $\lambda _\alpha \lambda _\beta $, which means that
the maximum number of supersymmetries is maintained. In all that
cases non-zero supercharges are neutral w.r.t. the algebra of
little group, and hence members of these supermultiplets are in
the same representation of generalized Poincare algebra. For other
configurations of branes situation  can be more complicated, as
discussed above. Particularly, the list of possible little groups
has to be derived. The rules for defining the statistics of
different states (unitary irreps of these little groups) can be
the following. First, one can try to define them directly. The
most relevant seems to be the derivation \cite{anast} of
spin-statistic relation in the framework of geometric quantization
method, which doesn't require features, absent in our case, such
as brane's quantum field theory. Instead,  a consideration of
Hamiltonian actions \cite{sour} of bosonic subalgebra of super
Poincare groups (\ref{1}), (\ref{2}), etc, is required. Second,
one can use reasonable assumption that statistics is not changing
under dimensional reduction. Of course, for the cases with usual
little groups, i.e. those for non-zero $P_\mu$ only, we can use
usual spin-statistics relation. Combination of  these approaches
should permit one to find all possible spin-statistics relations
for unitary irreps of generalized Poincare algebras.

\section{Acknowledgements}

This work is supported partially by INTAS grant \#99-1-590. I'm
indebted to R.Manvelyan for discussions.

\end{document}